# Online discussion forums for monitoring the need for targeted psychological health support: an observational case study of r/COVID19_support

Fathima Rushda Balabaskaran[1], Annabel Jones-Gammon[2], Rebecca How[2], Jennifer Cole[3]*

[1] Department of Computer Science, Royal Holloway University of London

[2] Geography Department, Royal Holloway University of London

[3] Department of Health Studies, Royal Holloway University of London

*corresponding author: Dr Jennifer Cole, Lecturer in Global and Planetary Health, Department of Health Studies, Royal Holloway University of London, Egham Hill, Egham, London TW20 0EX, United Kingdom

**Background:** The COVID-19 pandemic has placed a severe mental strain on people in general, and on young people in particular. Online support forums offer opportunities for peer-to-peer health support, which can ease pressure on professional and established volunteer services when demand is high. Such forums can also be used to monitor at-risk communities to identify concerns and causes of psychological stress. The aim of this study was to create and observe an online support forum to see if such a platform can be used for health surveillance or 'social listening', in this case to identify issues that are causing or exacerbating psychological stress and anxiety in university-age students.

**Methods:** We created and monitored r/COVID19_support, an online forum for people seeking support during the COVID-19 pandemic, on the platform Reddit. We set up the forum in February 2020, and observed and analysed posts made to it between February 2020 and July 2021. We identify posts made by users self-identifying as students or posting about college/university life, then coded these posts to identify emerging themes that related to triggers of psychological anxiety and distress.

**Results:** 147 posts were made to the forum by 111 unique users during the study period. A number of themes were identified by manual coding, included: feelings of grief associated with the loss of college-related life experiences, such as graduation ceremonies or proms; difficulties with focussing on online and self-guided learning; and fears for the future, in particular of graduating into a constrained job market. The identification of specific issues enabled users to be signposted to information to help them cope with address those particular concerns.

**Conclusion:** Monitoring peer-to-peer forums can help to identify specific issues with which vulnerable groups may require additional support, enabling users to be signposted on to high-quality information to address specific issues.

**GJMEDPH 2021; Vol. xx, issue x | OPEN ACCESS**

**Conflict of Interest**—none | **Funding— FB and AJG were funded under the** Summer Research and Education Placements Scheme of Royal Holloway, University of London. RH and JC received no funding for their participation.

## Introduction

Over the last two decades, many health and support communities have been raised on, or have migrated online [1], [2], meeting and organising in cyberspace rather than in more traditional settings. This has had particular value during the worldwide COVID19 pandemic, as whilst people can congregate in online spaces, they cannot carry the SARS-Cov2 virus with them: they do not have to worry about catching the virus or spreading it to others. There is, therefore, value in exploring the relationship between psychological health and online space, including what infrastructures, governance models, norms and culture best enable online spaces to support good health and improve mental well-being. Such spaces become, in and of themselves, a remedy for the psychosocial stresses of the pandemic.

In a public health emergency when health care systems may be overwhelmed, the ability of affected communities to support one another and share trusted, reliable information can have a significant impact on the extent to which the outbreak might be contained or its spread slowed. [3–7] Funk and Janssen [8] have modelled the impact of peer influence on disease spread. Cole [9] has investigated how reliable, trustworthy and timely public health information can be exchanged over peer-to-peer platforms a during health emergency, concluding that this is best achieved by using platforms that are already widely used by the affected population—in particular, the already existing and highly popular information sharing platform Reddit. In this paper, we explore the use of such forums not only for immediate peer-to-peer support but also as a way to undertake social listening [10] to identify concerns emerging within communities so that interventions to address these concerns can be targeted to the affected community, and to signal ‘information voids’ that are in danger of being filled with low-quality or deliberately misleading information [11]. Whilst social listening is widely employed to identify misinformation online [12], its use as a way of identifying additional support a community might need has not yet been widely explored.

### *Safe spaces during COVID19*

SARS-Cov2, the causative virus of the COVID19 pandemic was first recorded at the end of 2019 [13] in Wuhan, China but cases soon spread around the world. By the beginning of January 2022, more than 298 million cases had been reported worldwide, and 5.5 million deaths [14]. Alongside the infectious disease burden, a second epidemic of mental ill-health has been observed with short- and long-term impacts on depression, stress and anxiety disorders, and PTSD [15, 16]. In China, psychological assistance hotlines were set up in early February 2020 [17], in acknowledgement of the inherent dangers in face-to-face counselling and therapy sessions during the disease outbreak. The importance of providing emotional support and long-term consideration of psychological distress was swiftly acknowledged across the globe [18], for example in the UK [19] Brazil [20] and India [21]. However, a primary focus on the epidemiology, clinical features and management of COVID19 threatened to leave mental health provision under-resourced [22]. With traditional support structures over-stretched, online peer-to-peer support groups emerged, sometimes spontaneously, sometimes from within already existing communities, for those suffering from health anxiety either due to a pre-existing condition or emergent under the stress of COVID19 [23]. Whilst the potential for heightening health anxiety through engagement in forums where such issues are being discussed and foregrounded must not be overlooked, [24] as people may also share negative or worrying experiences that exacerbate feelings of fear and anxiety [25], forums have the potential to provide spaces of comfort and companionship [26].

### *Students under stress*

College and University-age students are particularly vulnerable to mental health issues, irrespective of the pandemic, due to pressures to perform exerted at a sensitive stage in their lives, whereby future actions and decisions will profoundly affect the future [28]. Studies from across the world [29], [30] that have measured students' mental health at different stages of pandemic lockdowns to compare results over time showed a marked increase in mental health challenges, with up to 91% reporting an overall negative impact on mental health since the beginning of the pandemic. Anxiety levels in a university in North Carolina, USA, increased from 18.1% before the pandemic to 25.3% within a period of just 4 months [30]. Saadeh et al, in a study from Jordan, reported rates of depression amongst students high as 71%, with a further 62.5% of students reporting that quarantine had an overall negative effect on their mental health [31]. Savage et al [28] argue that this puts students at risk during Covid-19 pandemic, in particular from relapses in those who have had problems before and towards the end of the academic year when exams increase academic pressure. This affected the 2020 graduation cohort more than the 2021 as they had time to adjust to such changes. Chaturvedi et al [29], showed that pressure on students from less privileged, disadvantaged backgrounds and/or dysfunctional families, have been at the greatest risk of developing mental health issues throughout the pandemic, due to living in environments less conducive thus affecting their ability to study affectively. Other factors identified as influencing student mental health include moving home, loss of income from off-campus employment, lack of access to academic advisors, and lack of interaction with other students [32].

These pressures have led to heightened issues of anxiety, depression, and stress in college-age students; manifesting in disturbed sleep patterns [31], [33] (which may be impacted by excessive screen time during quarantine as well as anxiety and stress [34]); an increase in negative expectations of the future and a loss of hope [27], leading to a declining spiral of depression; and limited movement and activity leading to more sedentary behaviour, which is also known to be a risk factor for depression [e.g. work by Savage et al, [28] on the UK during lockdown]. Uncertainty around how assessments and online learning were being organised [35] and graduations being cancelled [32] made students feel as though their college work was neither meaningful nor worthy. A study on 89.8% of Malaysian students reported family and work conflicts during the country's six-week lockdown in March-April 2020 [36] and in the US, 82% of students expressed increased concerns about academic performance during the pandemic [33].

***Creating safe space for support***

In this paper we explore how online forums might be used to ease some of these pressures, using as a case study r/COVID19_support, an online support forum we created at the beginning of the pandemic, hosted on the popular platform www.reddit.com. The forum operates as an online space that aims to create a supportive community meeting place and a location for exchanging knowledge that can help to ease anxiety; it thus offers a safe space in which people can discuss their fears, anxieties and bolster their psychological well-being, in line with explorations of concepts of space and place drawn from human geography [e.g. 30, 37, 38], organisational theory [e.g. 39–41] and computer science [e.g. 42–44].

Previous research has suggested that during disease outbreaks, people actively seek out others with knowledge and experience of similar events with whom to discuss their situation [9], [45], [46] in order to help them cope. They want to attend meetings, talk to experts, receive information from health professionals, and connect with friends and family who are experiencing the outbreak more immediately than themselves, for example those who are working in the healthcare sector, or in regions where the case counts are high. They want to learn more about the disease and the places in which it is circulating, particularly as it comes closer to them. Cyberspace can provide safe spaces in which conversations of this kind can happen and where knowledge can be both created and shared [41].

Whilst technology can increase social capital and agency, [47] attention has, also and rightly, been given to who is excluded from such spaces [48], [49], how access to technology creates digital divides between techno haves and have nots [50], and to the challenges of misinformation in uncontrolled information ecosystems [51], which can increase psychological distress as well as relieve it. Online surveillance during a pandemic may itself create concerns [52], particularly where government efforts such as contact tracing and movement tracking can be seen as excessive and restrictive of freedom [53]. These concerns are valid areas for study but the potential dangers should not distract from a full examination of what can happen inside safe digital spaces and what kind of health- and well-being enabling places can be formed there, particularly as there are studies that show the value of online mental health communication programmes as a coping strategy [54] and growing interest in the use of social media as an enabler of public health [55]. Our study highlights how social listening within online forums can identify specific concerns and thus enable targeted high-quality information to address those concerns to be targeted to those expressing them as well as enabling peer-to-peer support.

## Methods and materials

In January 2020, at the beginning of the COVID19 pandemic, our previous research on potential platforms for peer-to-peer communication during pandemics [9], [56] led us to consider setting up a forum on www.reddit.com, as a continuation of research into health information online undertaken during the 2014-15 West Africa Ebola crisis [9] and the 2018 Nipah virus outbreak [56]. Our previous research [9], [56], [57] has indicated that forums for high-quality peer-to-peer information exchange during disease outbreaks work well when existing popular and familiar platforms are used and the forum moderators have experience in the subject under discussion, the technical aspects of platform on which the discussion is hosted, and the norms and culture of the platform and its communities. At the beginning of the COVID19 pandemic, we chose to build a forum for peer-to-peer support on Reddit, a news aggregator site that also hosts discussion forums. It is one of the world's 20 most popular websites, allowing users to post content, comment on content, and vote on both posts and comments made by other users. Content is socially curated and promoted by site members through voting [58] characteristics our previous research has shown helps to keep highly upvoted content high-quality and trusted by other community members. Our previous research has shown that Reddit can host huge discussions and is capable of rapid growth [9]. Each subreddit is monitored and managed by volunteer moderators who are able to set rules and remove (manually and with the help of AI) posts and comments that are deemed inappropriate, offensive, or inaccurate (for example, those that contain information that is factually incorrect, use racist or sexist language or profanities, or are aggressive), resulting in an online space where useful and interesting information can be shared among users and where users can ask questions for which they have been able to find answers elsewhere. Further research has shown that while the quality of the information on such forums can be variable, on well-managed forums it can be scientifically accurate and in line with accepted medical practice [8] as when doctors were asked to rate the usefulness and accuracy of comments and posts, the ratings they gave the information validated the Reddit community's voting on which contained the better information. As such, we chose to build the forum on Reddit as it has many characteristics conducive to ensuring the promotion of reliable, trustworthy, and high-quality information, including its moderation, voting structure, complex system of trust markers [8], [9] and imbedded AI functions that help with the automatic removal of abusive, incorrect or otherwise problematic context.

Several forums dedicated to the discussion of the coronavirus pandemic had already emerged naturally at this point. We observed the activity on the existing forums and joined the moderators teams of several, including r/coronavirus, r/China_Flu and r/nCov. We observed, however, that users of these existing forums frequently expressed concerns about the virus and how it might impact them; they were often afraid or felt alone; they had questions they wanted to ask and often requested help and support with navigating information they could find complex and confusing. On the largest forum, r/coronavirus, however, many of these posts were removed in favour of linking only to professional media and scientific reports. When a user who identified as a qualified therapist (under the username u/thatreddittherapist), began to offer free online therapy to those expressing concerns in comments, we contacted her and suggested setting up a dedicated forum specifically for this purpose. Such a forum, www.reddit.com/r/COVID19_support, was set up (by JC, using the reddit user name u/JenniferColeRHUK) on 12 Feb 2020, as a space in which people could express their concerns, anxiety and fears over the COVID19 pandemic, share personal experiences and ask for advice and emotional support.

### *Recruiting a Moderator Team*

Once the subreddit had been initiated, we next needed to recruit an effective group of moderators to help keep the forum running. Our previous study [9] had identified three distinct moderation tasks that needed to be covered by a moderation team (or a single moderator): subject matter expertise, technical expertise, and experience in Reddit's norms and structure. We were able to access and draw on all of these through the Reddit community and construct an effective and efficient moderator team within a reasonably short time frame.

We set out to recruit a team of moderators that conformed to the skillset identified above and the group dynamic identified as most beneficial to enabling the emergence of collective intelligence [59], a form of crowd wisdom greater than the sum of its parts. The team we recruited included public health academics, a qualified counsellor, a medical doctor and a U.S. government biosafety officer. As the pandemic progressed, and a number of users were identified who consistently provided high-quality, scientifically accurate and emotionally supportive information to their peers. These were offered 'helpful contributor' flair – an identifier that showed next to their username and helped to signal to others that their advice was considered high-quality by the moderator team. The moderators themselves displayed flairs showing their qualifications or job title, e.g. PhD Global Health, or Verified Nurse.

The moderator team programmed AutoModerator, an inbuilt AI function that can automatically remove information posted by users that is insensitive (eg, racist language suggesting that lack of personal hygiene is responsible for the outbreak) or inaccurate (e.g., comments that suggest vaccines are completely ineffective against newly emerging strains of the disease) to ease the manual moderation burden.

Initially, the forum grew very quickly, experiencing a short period of exponential growth that coincided with the lead-in to the introduction of restrictive lockdowns in Europe and North America in mid-March 2020. The forum reached 1,000 subscribers on 28 Feb, 5,000 on 10 March, 10,000 on 17 March and 20,000 on 6 April. Following this, growth slowed but the forum continued to gain new subscribers steadily, reaching 30,000 on 18 October 2020. At the end of 2021, it had continued to grow to nearly 40,000 subscribers.

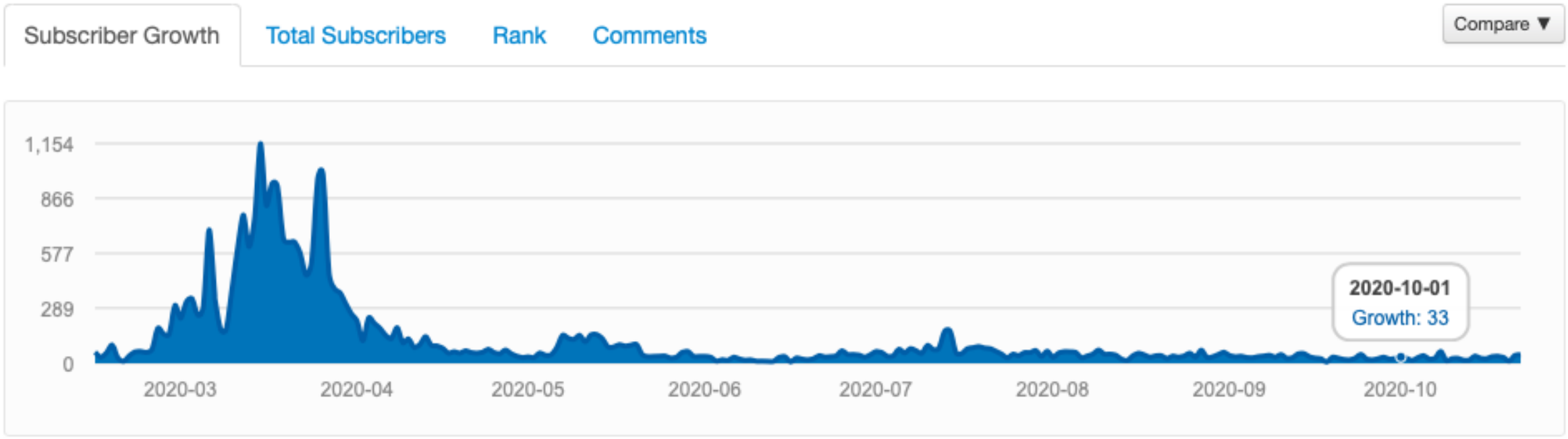

Fig 1: Subscriber growth of r/COVID19. New daily subscriptions peaked at 1154 on 15 March, with the main period of growth seen between 25 Feb and 4 April. Following that, the forum has grown steadily, gaining around 50 new subscribers per day. At the height of the pandemic in mid-2020 it regularly attracted between 500,000 and 1 million page views per day [60].

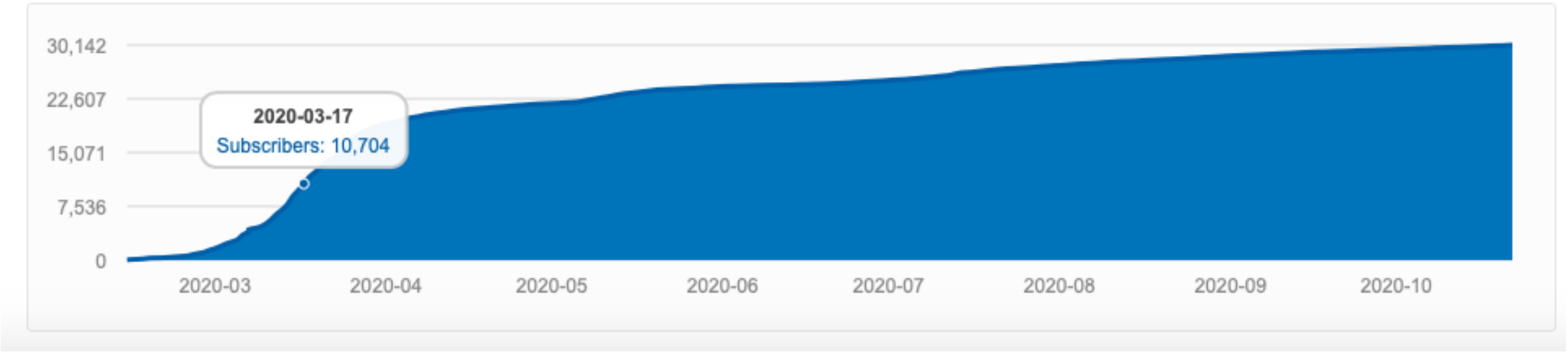

*Fig 2:* Cumulative growth of new subscribers on r/COVID19_support from February to October 2020, showing the exponential growth period during March 2020.

As the forum grew, additional moderators were added to meet the three key moderator skill sets identified in previous studies of Ebola forums: subject matter expertise (counselling skills, infection control, medical science, pandemic response, biosafety), experience of the platform (reddit, and previous disease outbreak forums or support and anxiety forums), and technical expertise of coding and platform architecture [9]. The active core team also met the criteria set out by Woolley [59] under which Collective Intelligence can most easily emerge in organisational teams: moderate diversity – of age, gender and geographic location – in a team of 6-8 people.

The decision to focus specifically on concerns raised by college and university age students emerged from a student placement undertaken by Rebecca How, one this paper's authors, to explore the extent to which the forum offered an opportunity to conduct social listening related to a specific health concern (mental health) within a specific demographic (college age students) and to design an intervention to address this through targeting appropriate information to this group.

A literature review was conducted to identify existing literature on student mental health during the COVID-19 pandemic using the search terms, 'Mental health of college students during covid- 19' and 'Mental health of university students during covid-19'. We then analysed the 200 most recent papers returned by each search on Google Scholar and removed duplicates. This returned 34 articles relevant to the study (Fig 1).

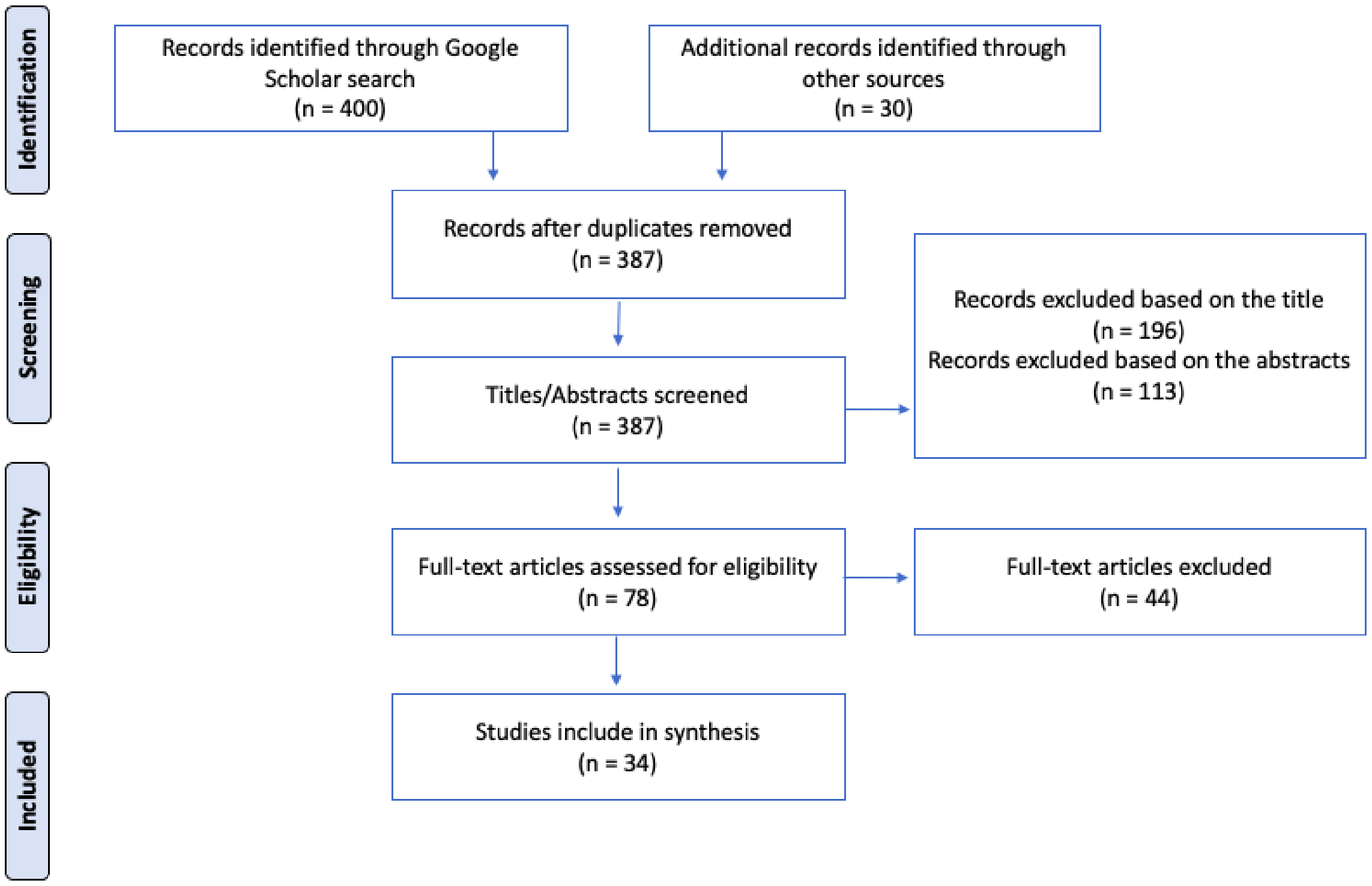

The selected 34 articles were identified by removing duplicate articles, then checking the titles to make sure that they were relevant to our research topic. We kept the titles that mentioned anything relating to the mental health of students during the COVID-19 pandemic. We then read the abstracts and excluded the articles that only briefly mentioned that one of the consequences of the pandemic is its impact on mental health but did not go into further detail. We then read the full articles and excluded any additional articles that did not have any sufficient information on how the mental health of students were affected or how the situation can be improved. Some papers contained similar content as they had referenced the same articles, and so these were further excluded, leaving the 34 articles that were included in the study.

### *Selecting posts for analysis*

Posts made to the forum by students from its inception up until May 31, 2021 were identified and analysed during a study period that ran from 1-31 May 2021. Posts were identified as being made by students if they were returned on a search for the keywords: 18th birthday, 21st birthday, class, college, freshman graduation, graduating, high school, homeschooling, masters, prom, school, semester, senior year, sophomore, student, teen, university. This returned 141 separate posts from 111 unique users. These posts were then coded for discussion themes. The posts were also analysed to identify which, if any, other reddit forums the posters engaged with; how similar posts were responded to on those other forums in comparison with on r/COVID19_Support; and whether posters seemed to create new accounts to post on the forum or used existing ones. Data was entered into an Excel Spreadsheet by RH, hand coded into topic areas by RH (checked by JC), and examined for links to other forums by RB, who also produced the visualisations in Fig 1 and 2.

## RESULTS

Fig 1 shows the number of posts returned for each keyword; the total is greater than the number of posts as some posts contained more than one of the keywords. The search returned 141 posts by 111 unique users. Most users (n=95, 85.5%) posted once only, 11 users (10%) posted twice, and five users (4.5%) posted three times. No user posted more than three times. Nearly half – 60/141, 43% of posts, were created by users who had created a new account in order to engage with the forum. Fig 1 shows the other forums on which r/COVID19_support users are posting, and Fig 2 groups these into broad topic areas, each of which will be considered in the discussion section.

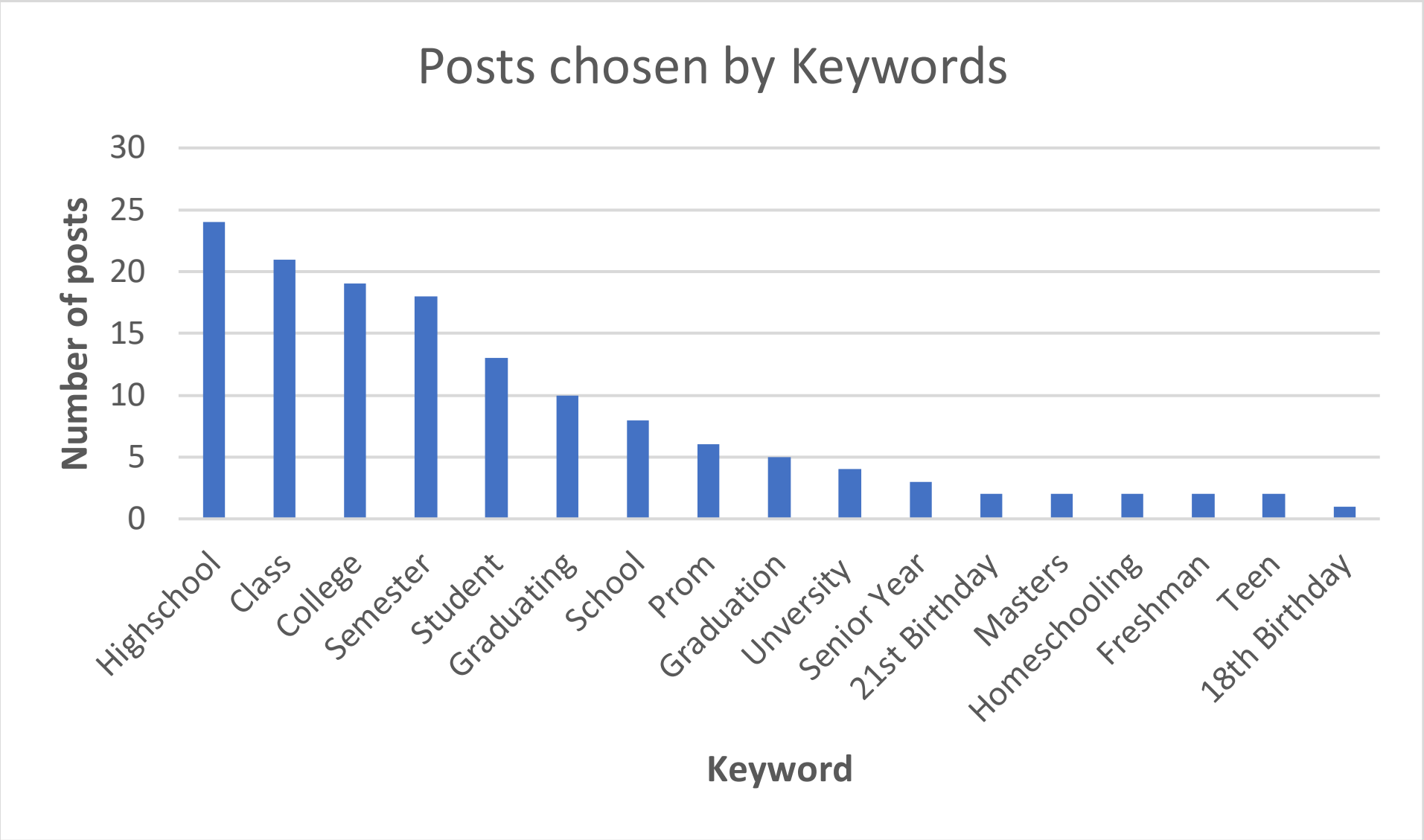

Fig 1: The number of posts returning one of the chosen keywords.

The content of the posts was analysed and initially grouped into categories by theme, of which the most commonly recurring ones where: challenges with online schooling (mentioned by n=31, 22% of posters); uncertainty for the future/disruption of plans such as taking a year out or moving out of the family home (n=40 + 8 + 5, 37.6%); missing or missing out on opportunities for social experiences (n= 16 + 24, 28.4%); general advice about COVID-19 and its risks that was not specifically related to student life (n=13, 9.2%); and unsupportive adults (teachers and parents); (n=3 + 1, 2.8%). The main subject areas of discussion this identified are shown in Fig 2, and more granular analysis of the content of the discussions is shown in Table 1.

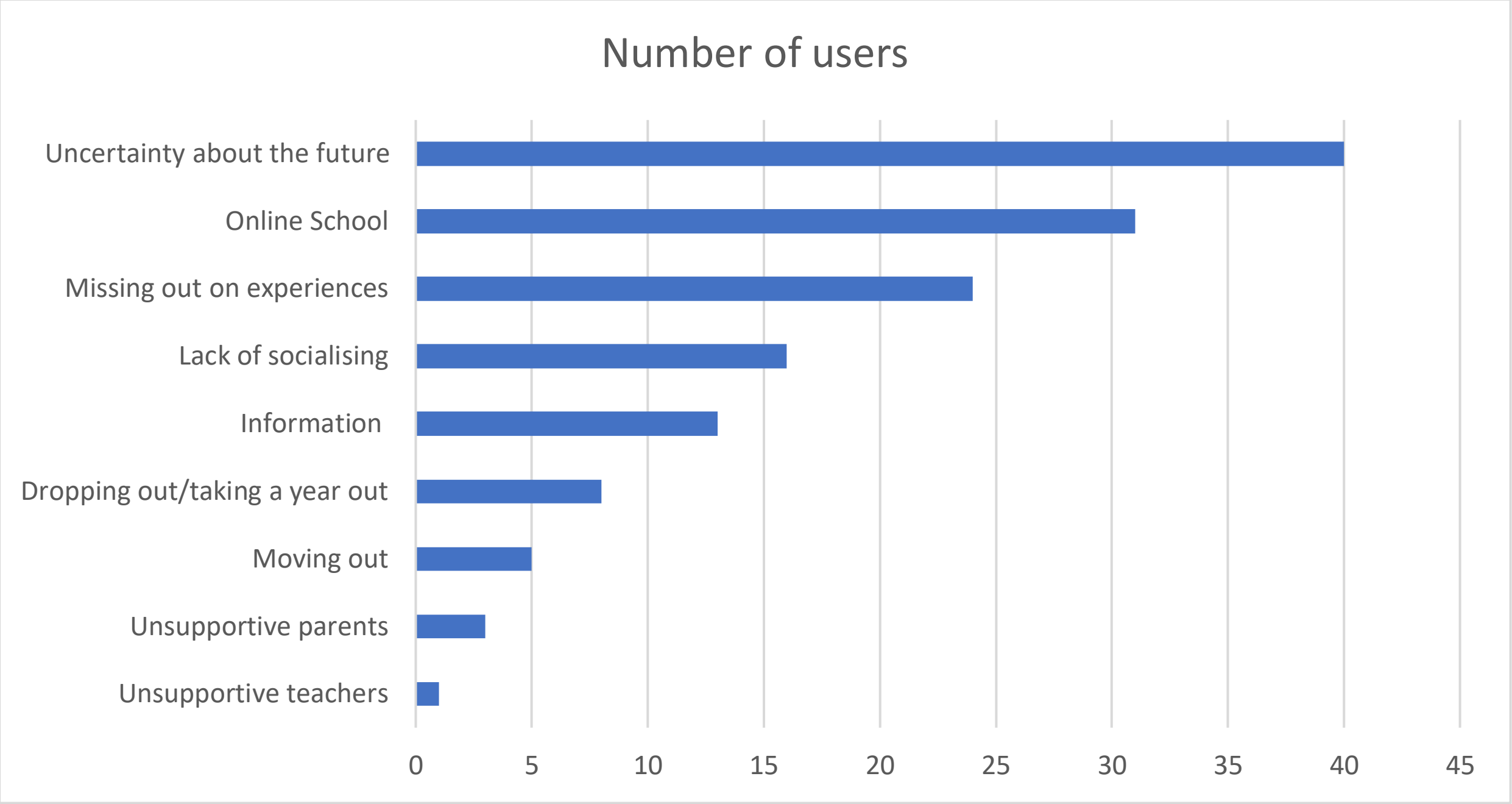

*Figure 2: Number of posts categorized by content.*

| | Theme | No | Specific issues |
|---|---|---|---|
| Online school | Online school | 31 | • Lack of motivation<br>• Homework and assignments overdue<br>• Failing classes<br>• Not able to concentrate<br>• Prefer in person teaching |
| Uncertainty about the future/disrupted plans | Uncertainty of the future | 40 | • Not able to find jobs<br>• Not sure if they're whole life will be wasted on covid (irrational thinking)<br>• Not sure if they can socialise in large groups anymore<br>• Stress that covid has caused permanent mental damage to them<br>• Not knowing when covid will end and life will go back to normal |
| | Dropping out/taking a year out | 8 | • Students who have already dropped out<br>• Students who have taken a year out<br>• Students who are asking for advice on if they should take a year out |
| | Moving out | 5 | • Having plans to move out disrupted due to covid<br>• Asking advice on if they should move out of the family home |
| Lost experience | Lack of social experiences | 16 | • Missing sports activities<br>• Missing friends, significant others, family<br>• Not being able to make new friends<br>• Losing friends due to lack of communication |
| | Missing out on college experiences | 24 | • Graduation, prom, sports events being cancelled<br>• Not being able to experience college/uni the way they should |
| General advice | General COVID info/advice | 13 | • General information about covid<br>• How long restriction will last<br>• How long they must wear masks for<br>• Safety of meeting with friends |
| Unsupportive adults | Unsupportive parents | 3 | • Parents who don't believe in mental health<br>• Parents who have contributed to making the user's mental health worse<br>• Parents who haven't been supportive<br>• Teenagers who don't have a good relationship with their parents |
| | Unsupportive teachers | 1 | • Not offering students mental health help<br>• Giving them more work than usual because they're at home<br>• Not offering any feedback for work |

*Table 1: Recurring themes in posts by those identified as college age students.*

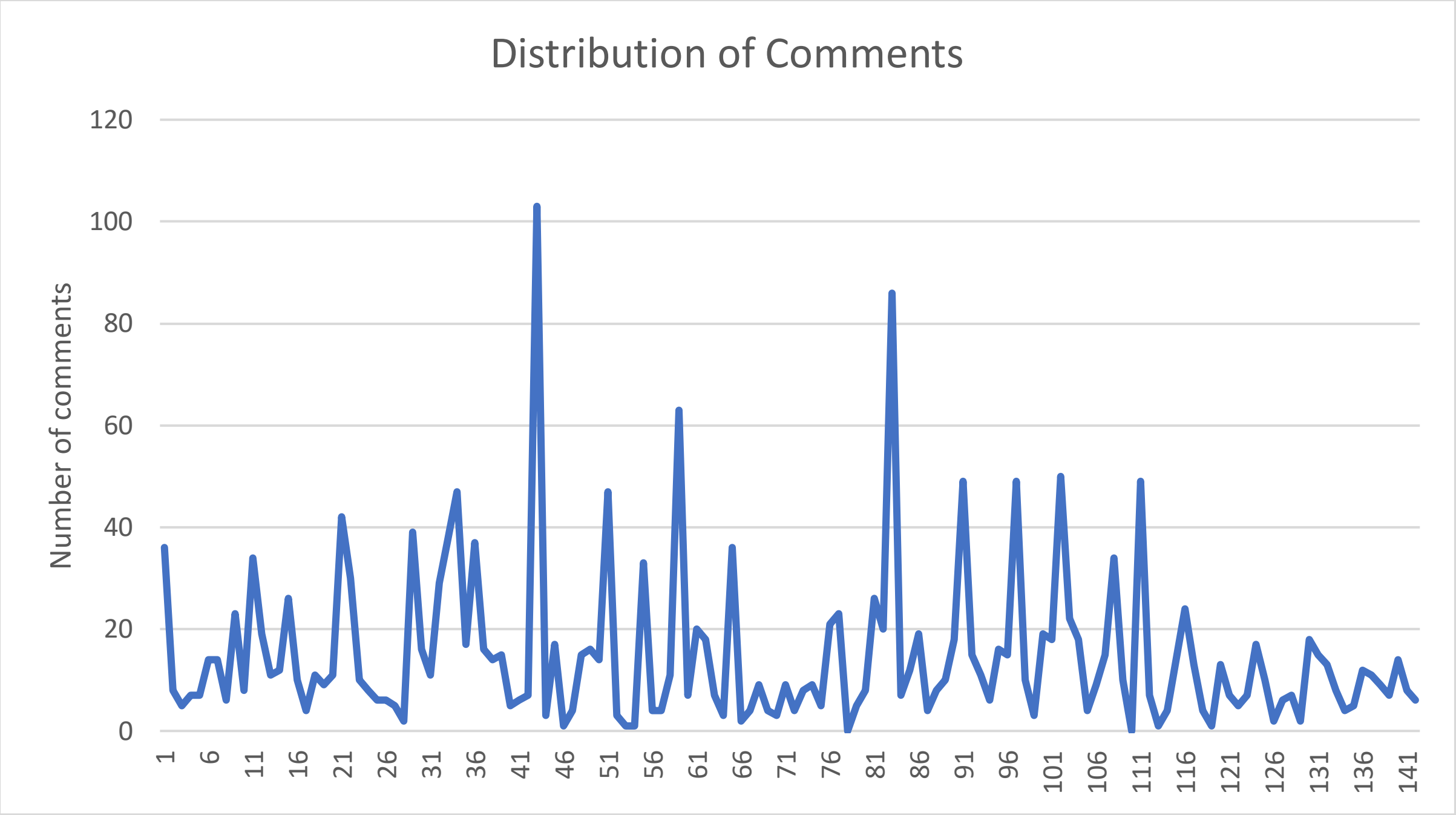

Fig 3: Average number of comments returned by posts.

The posts received on average 14.5 comments, with a range between 1–103 (Figure 3).

We then investigated what other reddit forums these users posted into and collected data on what they were posting, what they are looking for and how others react to their posts. There were 271 other forums that these users posted in, shown in Figure 4 below:

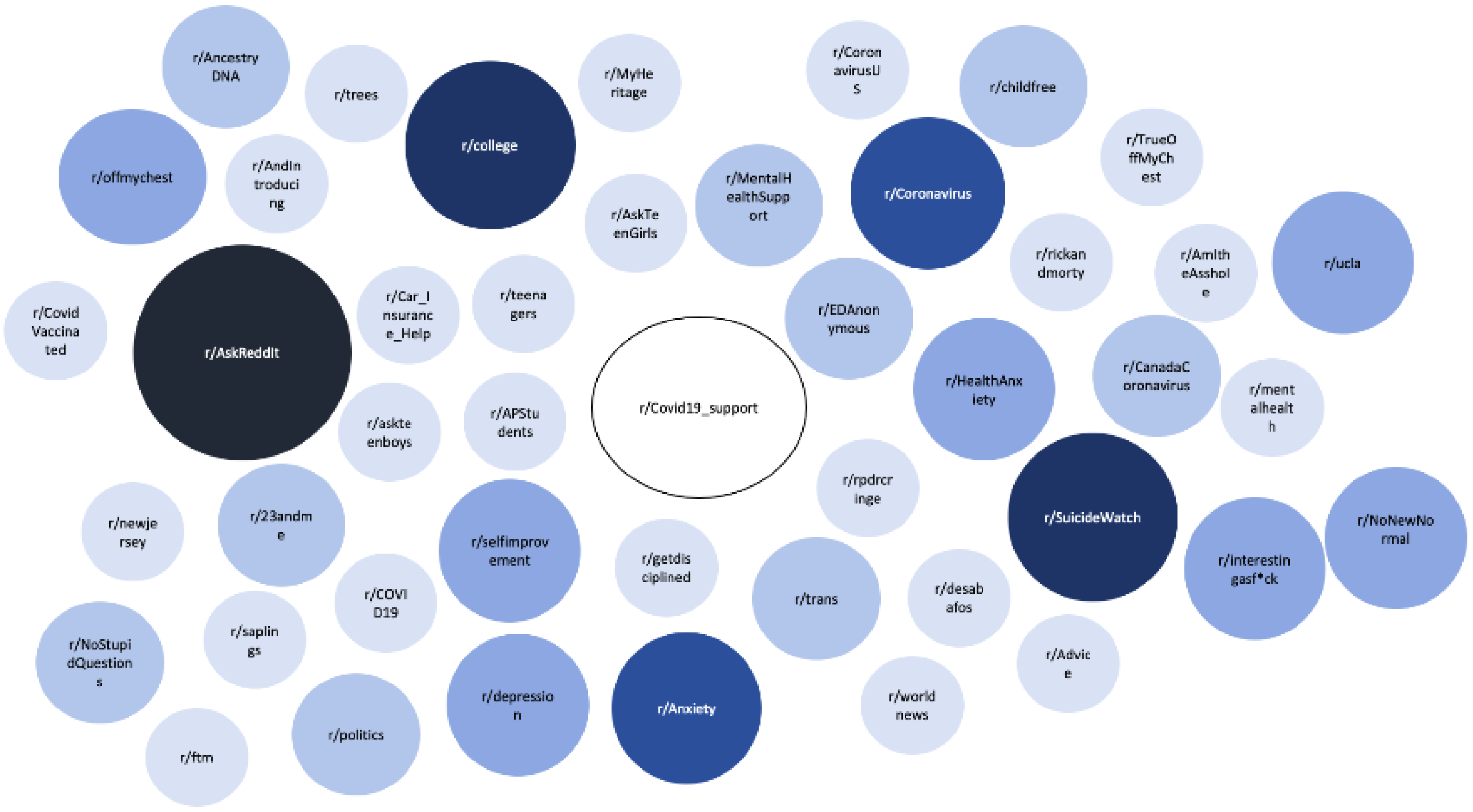

Figure 4: Other forusm on which r/COVID19_support users were posting. The side of the circle indicates the size of crossover between the fora.

These were then categorised these into the following groups based on the main topic of discussion in those forums: Hobbies (31.2%), Advice on non-COVID-19 issues (21.45%), School (13.93%), Mental Health (13.65%), other COVID-19 discussions (5.57%), Other general discussions (3.34%), Substance abuse/recreational drugs (3.34%), Family (2.3%), LGBTQ (1.95%); and Conspiracy (1.95%). Forums with considerably less than 2% crossover are not listed.

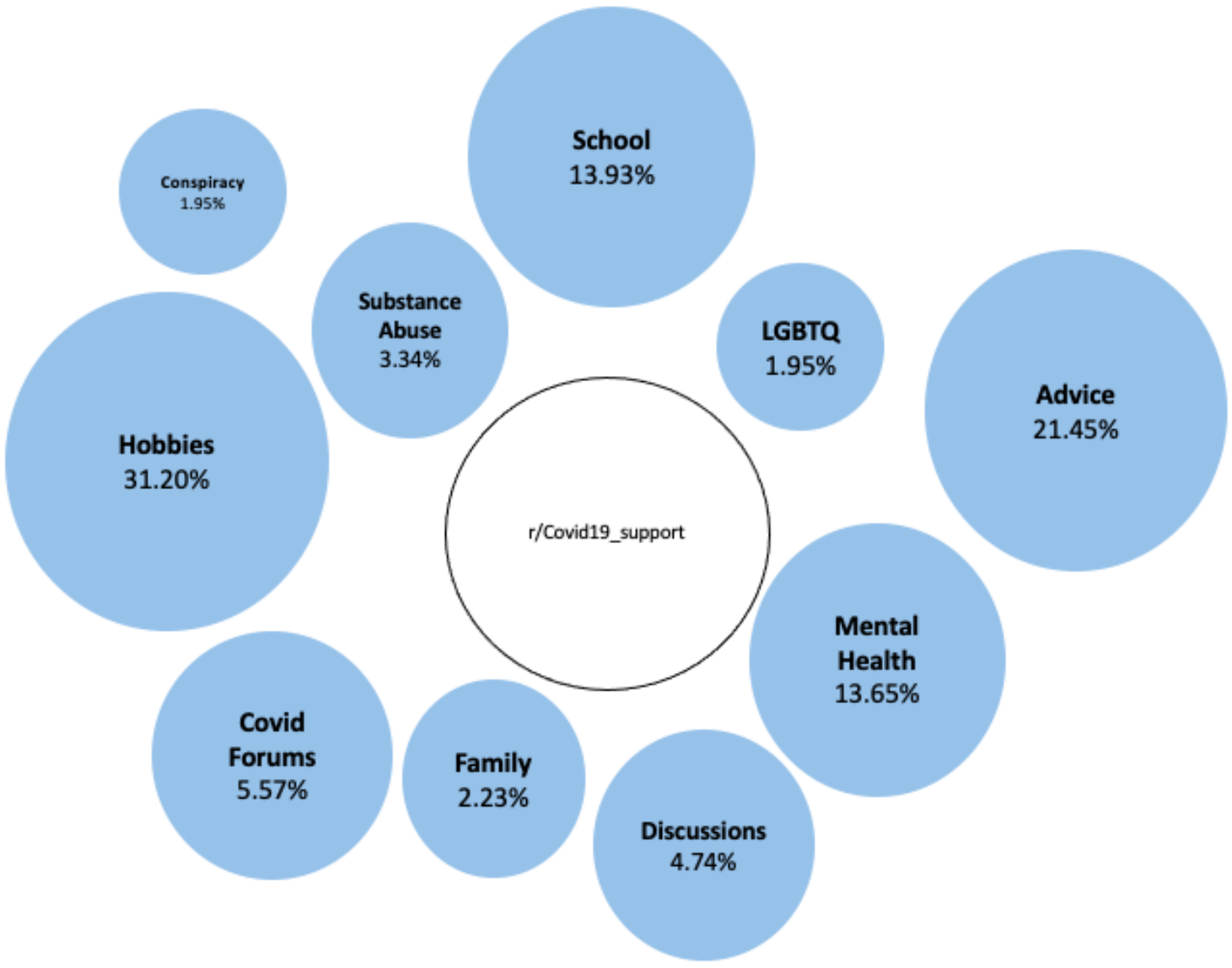

*Fig 5: Groupings showing the 10 most common grouping of other forum topics.*

## Discussion

From its inception, the r/COVID19_support forum attracted a steady stream of users, of whom several self-identified as students and young adults. These users came to the forum with concerns about the immediate impact the pandemic was having on their lives and also of what the long-term consequences might be. Their posts received responses that were most often supportive: other posters comforted them through sharing experience of similar issues, in particular their experiences of being a young adult around the time of the 11 September 2001 attacks on the US World Trade Centre or of graduating during the 2008-09 financial crash, and/or by giving useful and considerate advice (it is worth noting here that the forum is heavily moderated and unsupported comments are swiftly removed; many are caught by the automoderator AI function and never appear on the forum). Responders often referred them on to high-quality external content such as the support pages of national and international mental health charities or information pages of national and international public health agencies. Even comments that could be construed as negative (and sometimes were, by some users) were generally offering sound advice but was not necessarily what the poster most wanted to hear, e.g. that the current generation is not the only one to suffer and that things are probably not as bad as the students assumed.

Every post we analysed (n= 141, 100%) was from a user experiencing difficulty, anxiety or depression. This is not too surprising, as those who were coping well with the pandemic were unlikely to need to post on a support forum for those experiencing difficulties. It is worth noting, however, that 25 of the 142 users described pre-existing mental health problems which they felt had worsened due to the COVID-19 pandemic. In particular, the pandemic had been hard for students who had struggled socially at high school and felt that before the pandemic they had been beginning to come out of this as they embarked on college life. Sixteen of the posters felt progress in dealing with teenage mental health issues had been knocked back by the pandemic. The situation was particularly difficult for users who self-identified as autistic: six users mentioned that being autistic made online school difficult, including due to feeling overstimulated and more anxious. Students who were used to doing well in school but were finding themselves failing due to the difficulties of adjusting to online schooling were finding this experience particularly hard to cope with.

Users expressed that they well reassured when they realised others were facing similar issues, for example:

> "Thank you so much for your sentiment. Knowing that someone else is also in a similar situation makes me feel a little less alone."
>
> – u/smolcranium264 [61a]

> "I met many people on here that did not hesitate to help others just like I did and we ended up helping each other as well. While I may not know you guys I ending up seeing you as friends and it reminds me of the reason I want the pandemic to end."
>
> – u/alex_gaming_9987 [61b]

Users appreciated posts by older users who sought to assure them that while a year or two may seem like a lifetime when you are in your early twenties, it is not equivalent to one's entire life being wasted and that there will be life after the pandemic. This was particularly welcomed when specific examples were given, e.g. by users who had graduated during the 2008/9 financial crash but who had nonetheless gone on to have successful careers. Another common theme was posters feeling that they had not made the most of college/university or life in general before COVID-19 and had now lost the opportunity to do so; others were quick to reassure them that this was unlikely to be the case and that often life post-college/university could be more rewarding in any case.

A small number of posts (see Table 1) spoke of parents and teachers being unsupportive. Few comments offering advice to these posts suggested talking to parents about the challenges the users were facing, however. Peer-to-peer support seemed to be a preferred option, with many mentioning how grateful they were to have found r/COVID19_support and how much it had helped them.

This categorisation of poster concerns into a small number of broad categories enabled some of these to be addressed directly in a ‘Weekly Sticky’ [62], a post made by the moderators that is ‘stuck’ to the top of the forum’s front page and is thus the most prominent post to users entering the site. The ‘sticky’ offered some general advice on these themes and linked to useful external content to help users address issues and concerns, including general resources for maintaining mental health during the pandemic; tips on practicing mindfulness; resources for understanding the drivers of challenging emotions; advice on how to improve motivation and focus, particularly for online classes; how to cope with loneliness; how to understand and deal with feelings of loss and grief; and graduate anxiety. The post was 92% upvoted, indicating that it was well-received by the community.

Several studies have discussed the use of peer-to-peer online self-help forums similar to the Reddit COVID-19 support forum as a coping mechanism during the pandemic, including as a way for young people to connect with other users in similar situations [63], and to give advice and support. The app Sanvello provided by a US university as a wellbeing platform to talk to others in similar situations also supported students [33] and other studies have also found evidence of participants reaching out to website forums and mobile apps for support [64]. These studies did not, however, explore how such platforms were able to help those using them, nor if they had any specific characteristics or mechanism conducive to doing so. Listening to music, talking to friends and positive thinking have also proved to be popular approaches and led to fewer negative mental health impacts [65].

Many functions conducive to support are enabled by reddit’s platform architecture: some users posted videos of them performing songs they had written which they felt might help others, or shared online quiz nights. The forum thus acted as a community space as well as just a social media feed. There are clear links between the mental health of students and their ability to access and implement coping strategies [63], and the same study also found that those who had a high intolerance to uncertainty were more likely to seek out coping strategies. The implementation of online mental health communication programmes such as 'WeChat' provides an online psychological self-help prevention system to help anxiety and depression through cognitive behavioural therapy, for example [16] with benefits in an ability to alert volunteers to mentions of suicide or self-harm. This latter function is also embedded in the r/COVID19_Support forum, through the artificial intelligence programme Automoderator that can monitor and analyse messages posted, identify those at risk of suicide and alert the moderators who, in turn, can alert the platform administrators if it is felt escalation is needed.

Such online spaces are not, of course, without challenges. Quotes such as the one below identify some typical issues that users face:

> "I feel that social media might be one of my biggest triggers as of late with this pandemic. Every time I go online I see people in groups, living life, and everyone seems to be acting like we're back to normal. This sucks when my life hasn't changed at all since March 15th of last year" – u/jpark24 [66a]

> "I have a couple of family members and friends who […] ask 'what have you done with this lockdown?' It's so patronizing. Aunt Debora, I'm really happy that you've gotten into painting this year but I am holding on by a thread and I don't have the energy to get into pottery, so stop shaming me for it! – u/--MudKip-- [66b]

Five of the posts we analysed mention social media having a negative affect on their mental health, often because those who were following social distancing guidance and mask wearing frequently saw their peers not taking COVID-19 as seriously, and considered this unfair as they were sticking to the rules. One of the users mentioned their father had passing away after a colleague went to the beach and subsequently passed on COVID-19 onto him. The topic-specific forum of r/COVID-19_support offered some relief from this, however, as users could feel confident that other users shared their caution and preference for adhering to restrictions, and the forum also gave them an opportunity to vent about such experiences knowing they were likely to receive support and empathy from others.

Such forums may be particularly value to students as they are a demographic that has been found to be reluctant to seek help mental health support from professionals, despite their struggles [67] and so such peer-to-peer forums may provide a first-step engagement with mental health support. Whilst advice to avoid social media entirely has been suggested as a way of avoiding anxiety induced by pandemic-related bad news and concerns, this overlooks the supportive and unique environment such online spaces can offer to those who may be unable or unwilling to discussing their anxieties elsewhere or in person.

It was clear from our investigations that online forums can not only identify specific mental health challenges affecting a specific demographic, but may also identify where signposting to mental health support can be best-placed across platforms such as reddit that contain many distinct but intersecting communities. Our analysis of where else the students posted showed activity in forums for school and teen interests (e.g. r/teenagers, r/college and r/askteenboys) and in general advice forums (e.g. r/advice and r/nostupidquestions) which suggests that these might be places where mental health advice or links could be advertised and promoted help signpost those in need towards them. Also interesting to note is that there was a small but noticeable crossover with posting in r/trees and r/saplings, subreddits in which the recreational use of cannabis is discussed. Recreational use of an anxiety-using drug may well make users more susceptible to pandemic-related anxiety: clear signposting to where help can be found may be particularly valuable in such forums.

We recognise that there are, inevitably, limitations to our study. These include that we researched one forum only, for a relatively short space of time, and with regard to only one health event: the COVID-19 pandemic. A larger study, of several such forums would be required to build up a full understanding of the efficacy of such platforms. As users on reddit use pseudo-anonymised usernames, we were unable to collect demographic data such as country of origin or gender, and were reliant on assuming posters who self-identified as students were, indeed, from that demographic. We were not able, within the constraints of the project, to determine whether or not the advice offered to the posters would be considered appropriate by a mental health professional or counsellor; however, one of the moderator team is a professional counsellor and has no concerns over the general quality of advice; any that is considered inappropriate is removed from the forum. Despite these limitations, we do believe the study provides insights on how such forums might be used to monitor and support psychological well-being in young people during future events.

## Conclusions

We have previously theorized that social networking platforms such as Reddit can be a valuable knowledge exchange space in the event of an epidemic [8], [9], [45], [46]. New forums, where people can come together in the time of crisis, can be updated regularly with information regarding the ongoing event, and can receive advice on their concerns, can be set up quickly in the event of an unexpected public health emergency [56]. The COVID-19 pandemic has given us the opportunity to expand our understanding of the utility of such platforms by investigating their value for two specific health requirements: social listening to identify particular community issues and to provide psychological support for college-age students.

Our results suggest that Reddit, and platforms like it, can provide spaces in which health issues can be identified and discussed, and where information to address them can be shared safely. Such information can be distributed directly to communities likely to be particularly vulnerable, in this case, information on coping mechanisms for anxiety, loneliness, feelings of grief and worries for the future. Such forums may thus be able to ease pressure on professional services, monitor vulnerable communities to identify particular requirements needed to help them deal with challenging situations, and identify target audiences for the dissemination of information on how to deal with health anxiety during public health emergencies. They may also offer those less willing to engage with mental health support offline a first-step, anonymised opportunity to seek and receive help.

## Acknowledgments

This work would not have been possible without the funding provided by the Summer Research and Education Placements Scheme of Royal Holloway, University of London, which funded FB and AJG for their time. The project was started in the interest of JC to further interdisciplinary academic research. External funding may be required if we are to progress further with this topic of interest.

## Conflicts of Interest

None declared.